\documentclass[10pt, twocolumn, conference]{IEEEtran}
\usepackage[dvips]{graphicx}
\usepackage{amsmath}
\usepackage{amsthm}
\usepackage[flushleft]{threeparttable}
\usepackage{array}
\usepackage{booktabs}
\newcommand{\PreserveBackslash}[1]{\let\temp=\\#1\let\\=\temp}
\newcolumntype{C}[1]{>{\PreserveBackslash\centering}p{#1}}
\newcolumntype{R}[1]{>{\PreserveBackslash\raggedleft}p{#1}}
\newcolumntype{L}[1]{>{\PreserveBackslash\raggedright}p{#1}}

\usepackage{bm}
\usepackage{graphicx,subfigure}
\usepackage{algorithmic, algorithm}
\usepackage{cite}
\usepackage{amsmath}
\usepackage{amssymb}
\usepackage{psfrag}
\usepackage{empheq}
\usepackage{latexsym, amsmath, subfigure, color, amsfonts, amssymb,graphicx}
\usepackage{wrapfig} 
\newtheorem{Theorem}{Theorem}
\newtheorem{Lemma}{Lemma}

\newtheorem{Corollary}{Corollary}

\IEEEoverridecommandlockouts
\usepackage{tabularx}
\makeatletter
\def\hlinewd#1{%
\noalign{\ifnum0=`}\fi\hrule \@height #1 %
\futurelet\reserved@a\@xhline}
\makeatother
\begin{document}
\title{Centralized Coded Caching for Heterogeneous Lossy Requests}
\author{Qianqian Yang and Deniz G\"{u}nd\"{u}z\\
\IEEEauthorblockA{Dept. of Electrical and Electronic Eng., Imperial College London, UK\\
Email: \{q.yang14, d.gunduz\}@imperial.ac.uk}
}

\maketitle
\begin{abstract}
 \textit{Centralized coded caching} of popular contents is studied for users with heterogeneous distortion requirements, corresponding to diverse processing and display capabilities of mobile devices. Users' distortion requirements are assumed to be fixed and known, while their particular demands are revealed only after the \textit{placement phase}. Modeling each file in the database as an independent and identically distributed Gaussian vector, the minimum \textit{delivery rate} that can satisfy any demand combination within the corresponding distortion target is studied. The optimal delivery rate is characterized for the special case of two users and two files for any pair of distortion requirements. For the general setting with multiple users and files, a layered caching and delivery scheme, which exploits the successive refinability of Gaussian sources, is proposed. This scheme caches each content in multiple layers, and it is optimized by solving two subproblems: lossless caching of each layer with heterogeneous cache capacities, and allocation of available caches among layers. The delivery rate minimization problem for each layer is solved numerically, while two schemes, called the \textit{proportional cache allocation (PCA)} and \textit{ordered cache allocation (OCA)}, are proposed for cache allocation. These schemes are compared with each other and the cut-set bound through numerical simulations.
\end{abstract}
\section{Introduction}
Wireless data traffic is predicted to continue its exponential growth in the coming years, mainly driven by the proliferation of mobile devices with increased processing and display capabilities, and the explosion of available online contents. Current wireless architecture is widely acknowledged not to be sufficient to sustain this dramatic growth. A promising approach to alleviate the looming network congestion is to \textit{proactively} place popular contents, fully or partially, at the network edge during off-peak traffic periods (see, for example, \cite{golrezaei2012femtocaching, Gregori2015multi, maddah2014fundamental}, and references therein). 

Conventional caching schemes utilize orthogonal unicast transmissions, and benefit mainly from local duplication. On the other hand, by \textit{coded caching}, a novel caching mechanism introduced in\cite{maddah2014fundamental}, further gains can be obtained by creating multicasting opportunities even across different requests. This is achieved by jointly optimizing the \textit{placement} and \textit{delivery} phases. Coded caching has recently been investigated under various settings, e.g., decentralized coded caching~\cite{maddah2013decentralized}, online coded caching~\cite{pedarsani2014online}, distributed caching ~\cite{Caire2015distributing}, etc.

Most of the existing literature follow the model in \cite{maddah2014fundamental}, in the sense that each file is assumed to have a fixed size, and users are interested in the whole file. However, in many practical applications, particularly involving multimedia contents, files can be downloaded at various quality levels depending on the channel and traffic conditions, or device capabilities. This calls for the design of \textit{lossy} caching and delivery mechanisms.

We model the scenario in which each user has a preset distortion requirement known to the server. For example, a laptop may require high quality descriptions of requested files, whereas a mobile phone is satisfied with much lower resolution. Users may request any of the popular files, and the server is expected to satisfy all request combinations at their desired quality levels. We model the files in the server as independent sequences of Gaussian distributed random variables. Exploiting the successive refinability~\cite{equitz1991successive} of Gaussian sources, we derive the optimal caching scheme for the two-user, two-file scenario. For the general case, we propose an efficient coded caching scheme which considers multiple layers for each file, and first allocates the available cache capacity among these layers, and then solves the lossless caching problem with asymmetric cache capacities for each layer. We propose two algorithms for cache capacity allocation, namely \textit{proportional cache allocation (PCA)} and \textit{ordered cache allocation (OCA)}, and numerically compare the performance of the proposed layered caching scheme with the cut-set lower bound.

The most related work to this paper is~\cite{hassanzadeh2015distortion}, in which Hassanzadeh et al. solve the inverse of the problem studied here, and aim at minimizing the average distortion across users under constraints on the delivery rate as well as the cache capacities. In~\cite{timo2015rate}, authors also consider lossy caching taking into
account the correlation among the available contents, based on which the tradeoff between the compression rate, reconstruction distortion and cache capacity is characterized for single, and some special two-user scenarios.

The rest of the paper is organized as follows. We present the system model in Section~\ref{sec1}. Section~\ref{section:2} presents results on the case with two files and two users. General case is investigated in Section~\ref{sec2b}, including a lower bound on the delivery rate. Numerical simulations are presented in Section~\ref{sec4}.  Finally, we conclude the paper in Section VI.

\section{System Model}\label{sec1}

We consider a server that is connected to $K$ users through a shared, error-free link. The server has a database of $N$ independent files, $S_1$, ..., $S_N$, where file $S_i$ consists of $n$ independent and identically distributed (i.i.d) samples $S_{i,1}$, ..., $S_{i,n}$ from a Gaussian distribution with zero-mean and variance $\sigma^2$, i.e., $S_i \sim \mathcal{N}(0, \sigma^2)$, for $i=1, ..., N$.

The system operates in two phases. In the \emph{placement phase}, users' caches are filled with the knowledge of the number of users and each user's quality requirement; but without the particular user demands. Each user has a cache of size $M_kn$ bits, whose content at the end of the \emph{placement phase} is denoted by $Z_k$, $k=1, ..., K$. Users' requests, $\mathbf{d}\triangleq(d_1, ..., d_K)$, $d_k\in \{1, ..., N\}$, are revealed after the \emph{placement phase}. In the \emph{delivery phase}, the server transmits a single message $X^n_{(d_1, ..., d_K)}$ of size $nR$ bits over the shared link according to all the users' requests and the cache contents. Using $Z_k$ and $X^n_{(d_1, ..., d_K)}$, each user $k$ aims at reconstructing the file it requests within a certain distortion target $D_k$.

An $(n, M_1, ..., M_K, R)$ \textit{lossy caching code} consists of $K$ cache placement functions:
\[f^n_{k}: \underbrace{\mathbb{R}^n \times  \cdots \times \mathbb{R}^n}\limits_{N~files} \rightarrow \{1, ..., 2^{nM_k}\}~~\mbox{for}~~k=1, ..., K,\]
where $Z_k^n=f_k^n(S_1^n, ..., S_N^n)$; one delivery function:
\[g^n: \underbrace{\mathbb{R}^n \times  ... \times \mathbb{R}^n}\limits_{N~files} \times \underbrace{d_1 \times ... \times d_K}\limits_{K~requests}  \rightarrow \{1, ..., 2^{nR}\},\]
where $X^n_{(d_1, ..., d_K)}=g^n(S_1^n, ..., S_N^n, d_1, ..., d_K)$; and $K$ decoding functions:
\[h_k^n :  \{1, ..., N\}^K \times \{1, ..., 2^{nM_k}\} \times \{1, ..., 2^{nR}\} \rightarrow R^n,\]
where $\hat{S}_{k}^n = h_k^n(\mathbf{d}, Z_k^n, X^n)$. Note that each user knows the requests of all other users in the delivery phase.

We consider quadratic (squared-error) distortion, and assume that each user has a fixed distortion requirement $D_k$, $k=1, ..., K$. Without loss of generality, let $D_1 \geq D_2 \geq \cdots \geq D_K$. Accordingly, we say that a distortion tuple $\mathbf{D}\triangleq(D_1, ..., D_K)$ is \textit{achievable} if there exists a sequence of caching codes $(n, M_1, ..., M_K, R)$, such that
\[\lim_{n \rightarrow \infty} \frac{1}{n} \sum \limits_{j=1}^n (S_{d_k,j} - \hat{S}_{k,j})^2 \leq D_k, ~~~~ k=1, 2, ..., K,\]
holds for all possible request combinations $\mathbf{d}$. We reemphasize that $\mathbf{d}$ is not known during the \emph{placement phase}, while $\mathbf{D}$ is known. For a given distortion tuple $\mathbf{D}$, we define the \textit{cache capacity-delivery rate tradeoff} as follows:
\begin{align}\label{eq2}
R^\star(M_1, ..., M_K) \triangleq \inf\{R: \mathbf{D}~\mbox{is~achievable.}\}
\end{align}

Note that this problem is closely related to the classical rate-distortion problem. Let $R(D)$ denote the \textit{rate-distortion function} of a Gaussian source $S \sim \mathcal{N}(0, \sigma^2)$. We have $R(D) \triangleq \frac{1}{2}\log_2\frac{\sigma^2}{D}$~\cite{cover2012elements}.

In the sequel we heavily exploit the \textit{successive refinability} of a Gaussian source under squared-error distortion measure~\cite{equitz1991successive}. Successive refinement refers to compressing a sequence of source samples in multiple stages, such that the quality of reconstruction improves, i.e., distortion reduces, at every stage. A given source is said to be successively refinable under a given distortion measure if the single resolution distortion-rate function can be achieved at every stage. Successive refinement has been extensively studied in the source coding literature; please see~\cite{hassanzadeh2015distortion} for its use in the caching context.


\section{Optimal Lossy Caching: Two Users and Two Files $(N=K=2)$}\label{section:2}

In this section, we characterize the optimal cache capacity-delivery rate tradeoff for the lossy caching problem with two users ($K=2$) and two files ($N=2$). The target average distortion values for user 1 and user 2 are $D_1$ and $D_2$, respectively, with $D_1\geq D_2$. Let $r_1$ and $r_2$ be the minimum compression rates that achieve $D_1$ and $D_2$, respectively; that is $r_i \triangleq R(D_i) = \frac{1}{2}\log_2\frac{\sigma^2}{D_i}$, $i=1, 2$. This means that, to achieve the target distortion of $D_i$, the user has to receive a minimum of $nr_i$ bits corresponding to its desired file.
\begin{table*}
\label{table1}
\centering
\caption{Illustration of Cache Placement}
\begin{tabular}{|l|c|c|c|c|c|c|c|c|}
\hline
       & \multicolumn{6}{c|}{First Layer} & \multicolumn{2}{c|}{Second Layer} \\ \hline
$S_1$  & $A_1$    & $A_2$    & $A_3$   & $A_4$   & $A_5$   & $A_6$   & $A_7$                         & $A_8$                        \\ \hline
$S_2$  & $B_1$    & $B_2$    & $B_3$   & $B_4$   & $B_5$   & $B_6$   & $B_7$                         & $B_8$                         \\\hlinewd{1.2pt}
User 1 & $A_1\oplus B_1$    &          & $A_3, B_3$   &         & $A_5, B_5$   &         &                               &                              \\ \hline
User 2 &          & $A_2\oplus B_2$    &         & $A_4, B_4$   & $A_5, B_5$   &         & $A_7, B_7$                         &                              \\\hlinewd{1.2pt}
Case i  & $M_1$    & $M_2$    & $0$   & $0$   & $0$   & $r_1-M_1-M_2$   & $0$                         & $r_2-r_1$                        \\ \hline
Case ii  & $M_1$    & $r_1-M_1$    & $0$   & $0$   & $0$   & $0$   & $\frac{M_1+M_2-r_1}{2}$                         & $r_2-\frac{M_1+M_2-r_1}{2}$              \\ \hline
Case iii  & $r_1-l_1-2l_2$    & $0$    & $l_2$   & $l_2$   & $l_1$   & $0$   & $\min\{r_2-r_1, M_2/2\}$                         & $\max\{0, r_2-r_1-M_2/2\}$              \\ \hline
Case iv  & $0$    & $r_1-M_1$    & $M_1/2$   & $M_1/2$   & $0$   & $0$   & $r_2-r_1$                         & $0$              \\ \hline
Case v  & $0$    & $0$    & $0$   & $0$   & $r_1$   & $0$   & $r_2-r_1$                         & $0$              \\ \hline
\end{tabular}
\end{table*}
We first present Lemma 1 specifying the lower bound on the delivery rate for given $M_1$ and $M_2$ in this particular scenario, followed by the coded caching scheme achieving this lower bound. The proof of the lemma is skipped due to space limitations. 

\begin{Lemma}\label{lemma_NK2}
For the lossy caching problem with $N=K=2$, a lower bound on the cache capacity-delivery rate tradeoff is given by
\begin{align}
R^\star (M_1, M_2) \geq & R_c(M_1, M_2)= \max\{r_1-M_1/2,\nonumber \\ 
& ~~~~ r_2-M_2/2, r_1+r_2-(M_1+M_2),\nonumber\\ 
& ~~~r_1/2+r_2-(M_1+M_2)/2, 0\}~\mathrm{bpss}. \label{eq44}
\end{align}
\end{Lemma}
The first three terms in (\ref{eq44}) are derived from the cut-set lower bound, which will be presented for the general scenario in Theorem 1. 

Based on (\ref{eq44}), we consider five cases depending on the cache capacities of the users, illustrated in Fig.~1:
\begin{figure}
\label{fig1}
\centering
\includegraphics[width=0.8\linewidth]{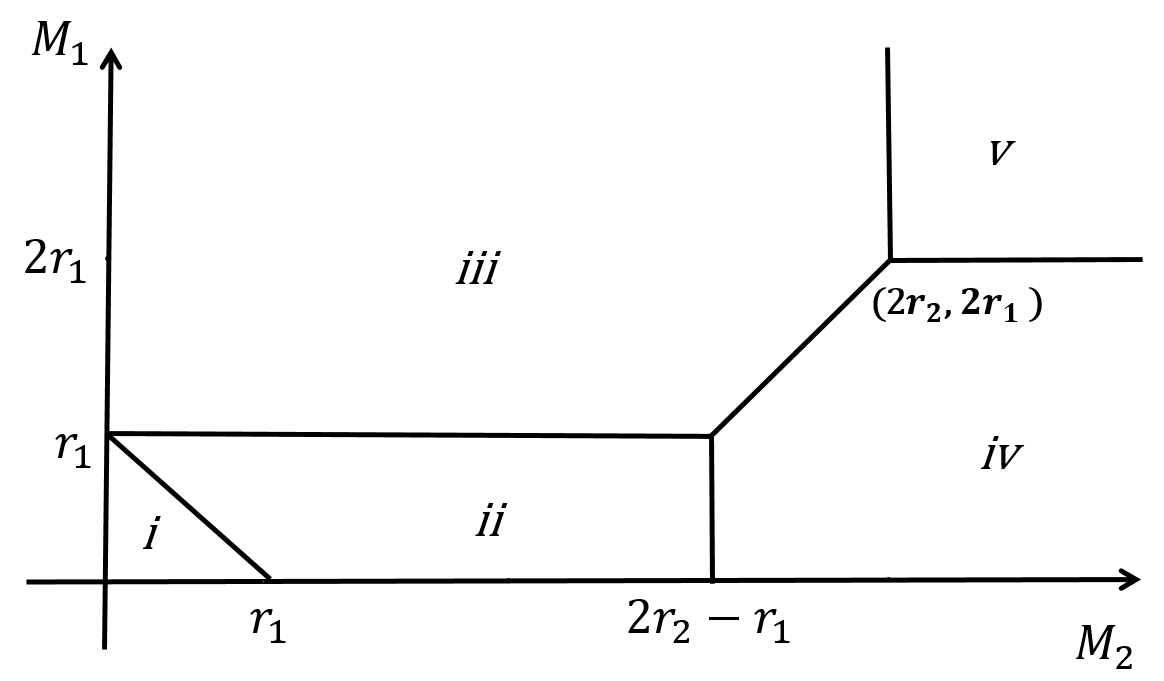}
\caption{Illustration of the five distinct cases of the cache capacities, $M_1$ and $M_2$, depending on the distortion requirements of the users, $r_1$ and $r_2$.}
\end{figure}

\textit{Case i}: $M_1 + M_2\leq r_1$. In this case, $R_c(M_1, M_2)=r_1+r_2-(M_1+M_2)~\mathrm{bpss}$.

\textit{Case ii}: $M_1 + M_2>r_1$, $M_1 \leq r_1$, $M_2 \leq 2r_2-r_1 $. We have $R_c(M_1, M_2)= \frac{r_1}{2}+r_2-\frac{M_1+M_2}{2}~\mathrm{bpss}$.

\textit{Case iii}: $M_1>r_1$, $M_2 \leq 2r_2$, $M_2-M_1 \leq 2r_2-2r_1$. Then $R_c(M_1, M_2)= r_2-\frac{M_2}{2}~\mathrm{bpss}$.

\textit{Case iv}: $M_1\leq 2r_1$, $M_2 > 2r_2-r_1$, $M_2-M_1 > 2r_2-2r_1$. It yields $R_c(M_1, M_2)=r_1-\frac{M_1}{2}~\mathrm{bpss}$.

\textit{Case v}: $M_1>2r_1$, $M_2>2r_2$. Then $R_c(M_1, M_2)=0$.

Next, for each of these cases, we explain the coded caching scheme that achieves the corresponding $R_c(M_1, M_2)$. We assume that the server employs an optimal successive refinement source code, denoted by $A(B)$ the source codeword of length $nr_2$ bits that can achieve a distortion of $D_2$ for file $S_1(S_2)$. Thanks to the successive refinability of Gaussian sources, a receiver having received only the first $nr_1$ of these bits can achieve a distortion of $D_1$. We refer to the first $nr_1$ bits as the first layer, and the $n(r_2-r_1)$ remaining bits as the second layer.

In each case, we divide the first layers of codewords $A$ and $B$ into six disjoint parts denoted by $A_1$, $\ldots$, $A_6$ and $B_1$, $\ldots$, $B_6$, and the second layers into two disjoint parts denoted by $A_7$, $A_8$ and $B_7$, $B_8$, such that $|A_i|=|B_i|$ for $i=1, ..., 8$, where $|X|$ denotes the length of the binary sequence $X$ (normalized by $n$).
%

Table~I illustrates the placement of contents in users' caches for each case. The second and third rows illustrate how the first and second layers are partitioned for each file. The fourth and fifth rows indicate the cache contents of each user at the end of the \emph{placement phase}. In all the cases, user 1 caches $Z_1=\{A_1\oplus B_1, A_3, B_3, A_5, B_5\}$ and user 2 caches $Z_2=\{A_2\oplus B_2, A_4, B_4, A_5, B_5, A_7, B_7\}$. The entries from the 6th row to the 10th specify the size of each portion in each case. For example, the 6th row implies that in \textit{Case i}, $|A_1|=|B_1|=M_1$, $|A_2|=|B_2|=M_2$,  $|A_6|=|B_6|=r_1-M_1-M_2$, $|A_8|=|B_8|=r_2-r_1$, and the sizes of all other portions are equal to $0$, which is equivalent to dividing $A(B)$ into four portions $A_1(B_1)$, $A_2(B_2)$, $A_6(B_6)$ and $A_8(B_8)$. Thus, in the placement phase, user 1 caches $Z_1=\{A_1\oplus B_1\}$, and user 2 caches $Z_2=\{A_2 \oplus B_2\}$ so that $|Z_1|=M_1$ and $|Z_2|=M_2$, which meets the cache capacity constraints. The cache placements of the other 4 cases are presented in a similar manner in Table~I.

Next, we focus on the delivery phase. We will explain the delivered message in each case to satisfy demands $\mathbf{d}=(S_1, S_2)$. All other requests can be satisfied similarly, without requiring higher delivery rates.

\textit{Case i} ($M_1 + M_2\leq r_1$): The server sends $B_1$, $A_2$, $A_6$, $B_6$ and $B_8$. Thus, the delivery rate is
\[R(M_1, M_2)= r_1+r_2-(M_1+M_2)~\mathrm{bpss}.\]

\textit{Case ii} ($M_1 + M_2>r_1$, $M_1 \leq r_1$, $M_2 \leq 2r_2-r_1 $): Server delivers $B_1$, $A_2$ and $B_8$. We have
\[R(M_1, M_2)= \frac{r_1}{2}+r_2-\frac{M_1+M_2}{2}~\mathrm{bpss}.\]

\textit{Case iii} ($M_1>r_1$, $M_2 \leq 2r_2$, $M_2-M_1 \leq 2r_2-2r_1$): The values of $l_1$ and $l_2$ in Table I are given as: $l_1=\max\{0,\min\{M_1-r_1, M_2/2-(r_2-r_1)\}\}$ and $l_2=\max\{0, M_2/2-(r_2-r_1)-l_1\}$. The server sends $B_1$, $B_3\oplus A_4$ and $B_8$ in the delivery phase, which results in
\[R(M_1, M_2)= r_2-\frac{M_2}{2}~\mathrm{bpss}.\]

\textit{Case iv}~($M_1\leq 2r_1$, $M_2 > 2r_2-r_1$, $M_2-M_1 > 2r_2-2r_1$): The server sends $B_2$, $B_3\oplus A_4$ and we have
\[R(M_1, M_2)=r_1-\frac{M_1}{2}~\mathrm{bpss}.\]

\textit{Case v} ($M_1>2r_1$, $M_2>2r_2$): The cache capacities of both users are sufficient to cache the required descriptions for both files. Thus, any request can be satisfied from local caches at desired distortion levels, and we have $R(M_1, M_2)=0$.

\begin{Corollary}
For $N=K=2$, the proposed caching scheme meets the lower bound in Lemma \ref{lemma_NK2}; and hence, it is optimal, i.e., we have $R^*(M_1, M_2)=R_c(M_1, M_2)$.
\end{Corollary}

\section{Lossy Coded Caching: General Case}\label{sec2b}

In this section, we tackle the lossy content caching problem in the general setting with $N$ files and $K$ users. Recall that the distortion requirements are assumed to be ordered as $D_1 \geq D_2 \geq \cdots \geq D_K$. Let $r_k=R(D_k)$, $k=1, ..., K$. Exploiting the successive refinability of Gaussian sequences, we consider a layered structure of descriptions for each file, where the first layer, called the $r_1$-description, consists of $nr_1$ bits, and achieves distortion $D_1$ when decoded. The $k$th layer, called the $(r_{k}-r_{k-1})$-refinement, $k=2, ..., K$, consists of $n(r_{k}-r_{k-1})$ bits, and having received the first $k$ layers, a user achieves a distortion of $D_k$. 

The example in Section~\ref{section:2} illustrates the complexity of the problem; we had five different cases even for two users and two files. The problem becomes intractable quickly with the increasing number of files and users. However, note that only users $k, k+1, ..., K$, whose distortion requirements are lower than $D_k$, need to decode the $k$th layer for the file they request, for $k=1, ..., K$. Therefore, once all the contents are compressed into $K$ layers based on the distortion requirements of the users employing an optimal successive refinement source code, we have, for each layer, a lossless caching problem. However, each user also has to decide how much of its cache capacity to allocate for each layer. Hence, the lossy caching problem is divided into two subproblems: the lossless caching problem of each source coding layer, and the cache allocation problem among different layers.

\subsection{Coded Lossless Caching of Each Layer}

Here we focus on the first subproblem, and investigate centralized lossless caching with heterogeneous cache sizes, which is unsolved in the literature, regarding each layer separately. Consider, for example, the $k$th refinement layers of all the files. There are only $L_k\triangleq K-k+1$ users (users $k, k+1, ..., K$) who may request these layers. Let user $j$, $j\in \{k, ..., K\}$, allocate $M_{j, k}$ (normalized by $n$) of its cache capacity for this layer. Without loss of generality, we order users $k, ..., K$ according to the cache capacity they allocate, and re-index them, such that $M_{k, k}\leq M_{k+1, k}\leq \cdots \leq M_{K, k}$.

We would like to have symmetry among allocated cache capacities to enable multicasting to a group of users. Based on this intuition, we further divide layer $k$ into $L_k$ sub-layers, and let each user in $\{k, ..., K\}$ allocate $M_{k}^1=M_{k, k}$ of its cache for the first sub-layer, and each user in  $\{k+i-1, ..., K\}$ allocate $M_{k}^i=M_{k+i-1, k}-M_{k+i-2, k}$ of its cache for the $i$th sublayer, for $i=2,\ldots, L_k$. Overall, we have $L_k$ sub-layers, and users $k+i-1, k+i, ..., K$ allocate $M_k^i$ of their caches for sub-layer $i$, whereas no cache is allocated by users $k, k+1, ..., k+i-2$.

We denote by $r_k^i$ the size of the $i$th sub-layer of the $k$th refinement layer, and by $R(L_k, i, M_{k}^i, r_k^i, N)$ the minimum required delivery rate for this sub-layer. The rates, $r_k^i$, $i=1, ..., L_k$, should be optimized jointly in order to minimize the total delivery rate for the $k$th layer. The optimization problem can be formulated as follows:

\begin{subequations}
\begin{equation}\min \limits_{r_k^1, ..., r_k^{L_k}} \sum_{i = 1}^{L_k} R(L_k, i, M_{k}^i, r_k^i, N)\end{equation}
\begin{equation}\mathrm{s. t.}  \sum_{i = 1}^{L_k} r_k^i=r_k-r_{k-1}.\end{equation}
\end{subequations}

We explore the achievable $R(L_k, i, M_{k}^i, r_k^i, N)$ based on the existing caching schemes in in~\cite{maddah2014fundamental} and \cite{chen2014fundamental}, which are referred to as \textit{coded delivery} and \textit{coded placement}, respectively. We consider two cases:

Case 1) $L_k < N$. In this case, coded placement scheme of \cite{chen2014fundamental} provides no global caching gain. Thus, we employ only coded delivery, and illustrate this scheme in our setup by focusing on the $i$th sub-layer: users $k+i-1$ to $K$ each allocate $M_{k}^i$ of cache capacity, while users $k$ to $k+i-2$ allocate no cache for this sublayer. If $r_k^i \in \{0, M_{k}^i/N, M_{k}^iL_k^i/((L_k^i-1)N), M_{k}^iL_k^i/((L_k^i-2)N), ..., M_{k}^iL_k^i/N\}$, where $L_k^i=L_k+1-i$,  we have
\begin{align} \label{eq3}
R(L_k, i, M_{k}^i, &r_k^i, N)=(i-1)\cdot r_k^i\\
&+r_k^iL_k^i\cdot(1- M_{k}^i/r_k^iN)\cdot\frac{1}{1+M_{k}^iL_k^i/r_k^iN}.\nonumber
\end{align}
The first term on the right hand side is due to unicasting to users $k$ to $k+i-2$, while the second term is the \textit{coded delivery} rate to users $k+i-1$ to $K$ given in ~\cite{maddah2014fundamental}. Based on the memory sharing argument, any point on the line connecting two points, $(r_1', R(L_k, i, M_{k}^i, r_1', N))$ and $(r_2', R(L_k, i, M_{k}^i, r_2', N))$, is also achievable, i.e., if $r_k^i \in [r_1', r_2']$, then we have
\begin{align} \label{eq4}
R(L_k, i, M_{k}^i, r_k^i , N)=&\frac{r_k^i-r_1'}{r_2'-r_1'}R(K_k, i, M_{k}^i, r_1', N)\nonumber\\
&+\frac{r_2'-r_k^i}{r_2'-r_1'}R(K_k, i, M_{k}^i, r_2', N),
\end{align}
where $r_1', r_2' \in \{0, M_{k}^i/N,M_{k}^iL_k^i/(L_k^i-1)N, M_{k}^iL_k^i/(L_k^i-2)N, ..., M_{k}^iL_k^i/N\}$; and if $r_k^i > M_{k}^iL_k^i/N$, we have
\begin{align} \label{eq5}
R(L_k, i, M_{k}^i, r_k^i, N)=&(i-1)\cdot r_k^i+\frac{M_{k}^iL_k(L_k-1)}{2N}\nonumber\\
&+(r_k^iM_{k}^iL_k/N)\times(L_k-i+1).\nonumber
\end{align}

Case 2) $L_k \geq  N$. In this case, \textit{coded placement} outperforms \textit{coded delivery} if the allocated cache capacity satisfies $M_k^i\leq\frac{r_k^i}{L_k^i}$\cite{chen2014fundamental}. Note that for the $i$th sub-layer, there are $i-1$ users with no cache allocation. If $i-1 \geq N$, there will be no gain with either schemes. When $i-1 < N$ and $r_k^i \geq M_{k}^iL_k^i$, the delivery rate of \textit{coded placement} is
\begin{equation} \label{eq6}
R(L_k, i, M_{k}^i, r_k^i, N)=Nr_k^i-M_{k}^ir_k^i(N-i+1).
\end{equation}
When $0\leq r_k^i\leq M_{k}^iL_k^i$, the delivery rate is given by the lower convex envelope of points $(M_{k}^iL_k^i, R(L_k, i, M_{k}^i, M_{k}^iL_k^i, N))$ given by (\ref{eq6}) and $(r_k^i, R(L_k, i, M_{k}^i, r_k^i, N))$, and for $r_k^i \in \{0, M_{k}^i/N,M_{k}^iL_k^i/((L_k^i-1)N), M_{k}^iL_k^i/((L_k^i-2)N), ..., M_{k}^iL_k^i/N\}$, given by (\ref{eq3}).

\subsection{Allocation of Cache Capacity}

We propose two algorithms for cache allocation among layers: \textit{proportional cache allocation} (PCA) and \textit{ordered cache allocation} (OCA), which are elaborated in Algorithms~1 and~2, respectively, where $r_k$ is as defined earlier, and we let $r_0=0$. .

\begin{algorithm}[htbp]\label{alg3}
\caption{Proportional Cache Allocation (PCA)}
\begin{algorithmic}
\STATE{
\textbf{Require:} $\mathbf{r}={r_1, ..., r_K}$
 \begin{enumerate}
\item for all $k \in {1, ..., K}$
\item ~~~~~for all $i \in {1, ..., k}$
\item ~~~~~~~user $k$ allocates $\frac{r_i-r_{i-1}}{r_k}M_k$ to layer $i$
\item ~~~~~end for
\item end for
\end{enumerate}
   }
\end{algorithmic}
\end{algorithm}

\begin{algorithm}[htbp]\label{alg4}
\caption{Ordered Cache Allocation (OCA)}
\begin{algorithmic}
\STATE{
\textbf{Require:} $\mathbf{r}={r_1, ..., r_K}$
 \begin{enumerate}
\item for all $k \in {1, ..., K}$
\item ~~~~~~~user $k$ allocates all of its cache to the first $i$ layers, where $r_{i-1} < \frac{M_k}{N} \leq r_{i}$
\item end for
\end{enumerate}
   }
\end{algorithmic}
\end{algorithm}
PCA allocates each user's cache among the layers it may request proportionally to the sizes of the layers, while OCA gives priority to lower layers. The server can choose the one resulting in a lower delivery rate. Numerical comparison of these two allocation schemes will be presented in Section V.

\subsection{Lower Bound}

The following lower bound is obtained using cut-set arguments.
\begin{Theorem}(Cut-set Bound)
For the lossy caching problem described in Section~\ref{sec1}, the optimal achievable delivery rate is lower bounded by
\begin{align}
\operatorname*{max}\limits_{s\in \{1,...,K\}} \operatorname*{max}\limits_{\begin{subarray}{c}
    \mathcal{U} \subset \{1,...,K\}\\
    |\mathcal{U}|=s
  \end{subarray}}\left(\sum\limits_{k\in \mathcal{U}} r_{k}-\frac{\sum\limits_{k\in \mathcal{U}}M_k}{\lfloor N/s\rfloor}\right). \nonumber
\end{align}
\end{Theorem}
\section{Simulations}\label{sec4}
In this section, we numerically compare the achievable delivery rates for uncoded caching, the proposed caching schemes, and the lower bound. In Fig. 2, we consider $K=10$ users and $N=10$ files in the server. Cache sizes of the users are identical, i.e., $M_1=M_2=\cdots=M_{10}=M$. The distortion levels $(D_1, D_2, ..., D_{10})$ are such that $(r_1, r_2, ..., r_{10})= (1, 2, ..., 10)$. While we observe that the proposed coded caching scheme greatly reduces the delivery rate, OCA performs better for small cache sizes, while PCA dominates as $M$ increases. Using memory sharing, we can argue that the dotted curve in Fig. 2, which is obtained through the convex combination of the delivery rates achieved by the two proposed schemes, is also achievable.
\begin{figure}\label{fig4}
\centering
\includegraphics[width=0.895\linewidth]{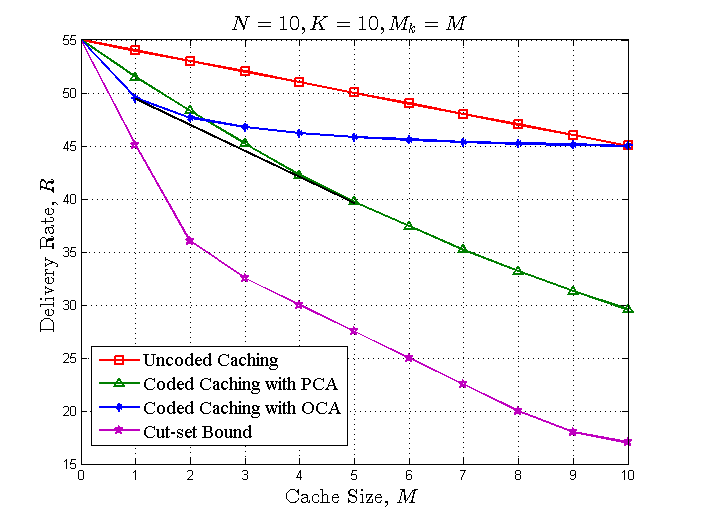}
\caption{Delivery rate vs. cache capacity with identical cache sizes.}
\end{figure}

In Fig. 3, we consider the same setting but with heterogeneous cache sizes,  where $M_k=0.2kM$, for $k=1, ..., 10$. In this setting, PCA allocates the same amount of cache to each layer at different users, which creates symmetry among the caches. The achievable delivery rates in Fig. 3 illustrate significant improvements in coded caching with PCA over both uncoded and OCA schemes in terms of the achievable delivery rates. We observe that the gains become more significant as the cache capacity, $M$, increases. While the lower bound is not tight in general, we see in both figures that the PCA performance follows the lower bound with an approximately constant gap over the range of $M$ values considered.


\begin{figure}\label{fig5}
\centering
\includegraphics[width=0.92\linewidth]{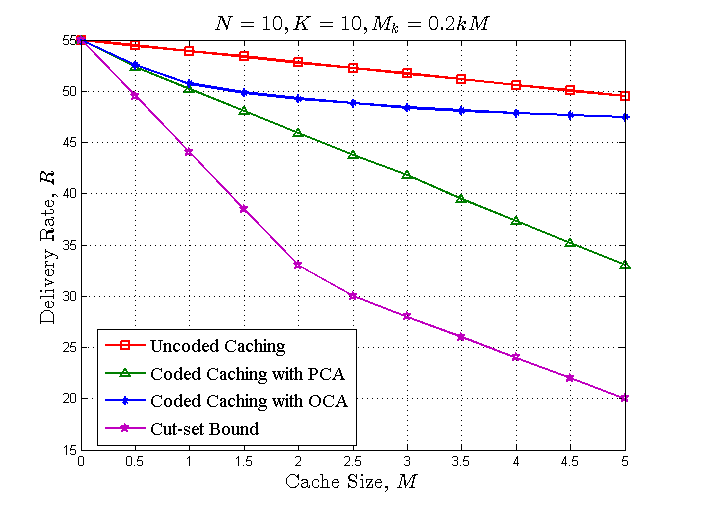}
\caption{Delivery rate vs. cache capacity with heterogeneous cache sizes.}
\end{figure}

\section{Conclusion}\label{sec5}

We investigated the lossy caching problem where users have different distortion requirements for the reconstruction of contents they request. We proposed a coded caching scheme that achieves the information-theoretic lower bound for the special case with two users and two files. Then, we tackled the general case with $K$ users and $N$ files in two steps: delivery rate minimization, which finds the minimum delivery rate for each layer separately, and cache allocation among layers. We proposed two different algorithms for the latter, namely, PCA and OCA. Our simulation results have shown that the proposed PCA scheme improves the required delivery rate significantly for a wide range of cache capacities; and particularly when the users' cache capacities are heterogenous. 
\bibliographystyle{unsrt}

\begin{thebibliography}{10}

\bibitem{golrezaei2012femtocaching}
 N.~Golrezaei, K.~Shanmugam, A. G.~Dimakis, A.~F.~Molisch and G.~Caire,
``Femtocaching: wireless video content delivery through distributed caching helpers,''
\newblock in {\em Proc. IEEE INFOCOM}, Orlando, FL, Mar. 2012, pp.1107--1115.

\bibitem{Gregori2015multi}
M.~Gregori, J.~Gomez-Vilardebo, J.~Matamoros and  D.~G\"{u}nd\"{u}z,
``Wireless content caching for small cell and {D}2{D} networks,''
\newblock {\em to appear, IEEE J. Sel. Areas Commun.}, 2016.

\bibitem{maddah2014fundamental}
M.~Maddah-Ali and  U.~Niesen, ``Fundamental limits of caching,''
\newblock {\em IEEE Trans. Inform. Theory}, vol. 60, no. 5, pp. 2856-2867, May 2014.

\bibitem{chen2014fundamental}
Z.~Chen, P.~Fan and K.~B.~Letaief,  ``Fundamental limits of caching: Improved bounds for small buffer users,''
\newblock {\em ArXiv:1407.1935v2 cs.IT}, Nov. 2015.

\bibitem{maddah2013decentralized}
M.~A.~Maddah-Ali and U.~Niesen, ``Decentralized coded caching attains order-optimal memory-rate tradeoff,''
\newblock {\em IEEE/ACM Trans. Netw}, vol. 23, no. 4, pp. 1029-1040 Apr. 2014.

\bibitem{pedarsani2014online}
 R.~Pedarsani, M.~Maddah-Ali and U.~Niesen, ``Online coded caching,''
\newblock {\em ArXiv:1311.3646 cs.IT}, Nov. 2013.

\bibitem{Caire2015distributing}
M.~Ji, G.~Caire and A.~F.~Molisch, ``Fundamental limits of distributed caching in D2D wireless networks,''
\newblock in {\em Proc. IEEE Inform. Theory Workshop (ITW)}, Jeju Island, Korea, Oct. 2015, pp. 1--5.

\bibitem{hassanzadeh2015distortion}
P.~Hassanzadeh, E.~Erkip, J.~Llorca and A.~Tulino, ``Distortion-memory tradeoffs in cache-aided wireless video delivery,''
\newblock {\em ArXiv:1511.03932 cs.IT}, Nov. 2015.

\bibitem{timo2015rate}
R.~Timo, S.~S.~Bidokthi, M.~Wigger and B.~Geiger, ``A rate-distortion approach to caching,''
\newblock in {\em Proc. Int'l. Zurich Seminar (IZS)}, Zurich, Switzerland, Mar. 2016.

\bibitem{cover2012elements}
T.~M.~Cover and J.~A.~Thomas,
\newblock {\em Elements of Information Theory}, John Wiley \& Sons, 2012.

\bibitem{equitz1991successive}
T.~M.~Cover and W.~H.~Equitz, ``Successive refinement of information,''
\newblock {\em IEEE Trans. Inform. Theory}, vol. 37, no. 2, pp. 269--275 , Mar. 1991.

\end{thebibliography}

\end{document}